\pdfoutput=1
\documentclass[prl,superscriptaddress,twocolumn]{revtex4}
\usepackage{epsfig}
\usepackage{bm}

\newcommand{\Wcm}{\;\mathrm{W/cm}^2}
\newcommand{\IP}{I_p}
\newcommand{\vk}{\mathbf{k}}
\newcommand{\vT}{\mathbf{T}}
\newcommand{\vE}{\mathbf{E}}
\newcommand{\ver}{\mathbf{r}}
\newcommand{\vA}{\mathbf{A}}
\newcommand{\vK}{\mathbf{K}}
\newcommand{\vd}{\mathbf{d}}
\newcommand{\ve}{\hat{\mathbf{e}}}
\newcommand{\veK}{\hat{\mathbf{e}}_\mathbf{K}}

\begin{document}

\title{High-order harmonic generation in polyatomic molecules induced by a bicircular laser field}
\author{S. Od\v zak}
\affiliation{Faculty of Science, University of Sarajevo, Zmaja od
Bosne 35, 71000 Sarajevo, Bosnia and Herzegovina}
\author{E. Hasovi\'{c}}
\affiliation{Faculty of Science, University of Sarajevo, Zmaja od
Bosne 35, 71000 Sarajevo, Bosnia and Herzegovina}
\author{D. B. Milo\v{s}evi\'{c}}
\affiliation{Faculty of Science, University of Sarajevo, Zmaja od
Bosne 35, 71000 Sarajevo, Bosnia and Herzegovina}
\affiliation{Max-Born-Institut, Max-Born-Str.~2a, 12489 Berlin,
Germany}
\affiliation{Academy of Sciences and Arts of Bosnia and
Herzegovina, Bistrik 7, 71000 Sarajevo, Bosnia and Herzegovina}
\date{\today}
\begin{abstract}
High-order harmonic generation by a bicircular field, which consists
of two coplanar counter-rotating circularly polarized fields of
frequency $r\omega$ and $s\omega$ ($r$ and $s$ are integers), is
investigated for a polyatomic molecule. This field possesses
dynamical symmetry, which can be adapted to the symmetry of the
molecular Hamiltonian and used to investigate the molecular
symmetry. For polyatomic molecules having the $C_{r+s}$ symmetry
only the harmonics $n=q(r+s)\pm r$, $q=1,2,\ldots$, are emitted
having the ellipticity $\varepsilon_n=\pm 1$. We illustrate this
using the example of the planar molecules BH$_3$ and BF$_3$, which
obey the $C_3$ symmetry. We show that for the BF$_3$ molecule,
similarly to atoms with a  $p$ ground state, there is a strong
asymmetry in the emission of high harmonics with opposite
helicities. This asymmetry depends on the molecular orientation.
\end{abstract}


\maketitle

Strong-field physics and attoscience are presently a very active
area of research since they enable one to study matter on the scale
of the electronic dynamics \cite{KrauszRMP}. The electron liberated
by a strong linearly polarized laser field can return to the parent
ion and trigger various strong-field processes, which include
elastic rescattering (the process in which the final electron energy
is much higher than in direct ionization is called high-order
above-threshold ionization \cite{adv12}), the inelastic process in
which in an ionizing collision one or more additional electrons are
liberated \cite{RMPNSDI}, and the laser-assisted electron-ion
recombination process where a high-energy photon is emitted [this is
the so-called high-order harmonic generation (HHG)
\cite{Kohler2012}]. The dynamics of electrons driven by a linearly
polarized laser field is one-dimensional and with it it is difficult
to explore structure and dynamics of more complex targets such as
molecules (for reviews about molecular strong-field processes see
\cite{LeinJPB2007,lin-review}). Since symmetry is a fundamental
concept of science and since it is the key for understanding the
structure and dynamics of molecules, it is important to find field
configurations that possess both particular symmetry properties and
the possibility that the laser-field-driven liberated electron
returns to the parent ion. Such a field will be the main subject of
the present study.

The concept of symmetry is particularly relevant for polyatomic
molecules. Due to the complexity of polyatomic molecules, research of
molecular HHG has mainly been confined to simple diatomic and
triatomic linear molecules. A list of recent papers in
which polyatomic molecules were studied includes
\cite{WagnerPNAS2006,TorresPRL2007,LiScience2008,WongPRA2010,WongPRA2011,VozziNP2011,WongPRL2013,LePRA2013,FerreNC2015,KrausNC2015,Manschwetus2015,SandorPRL2016}.
Ab initio HHG calculations for polyatomic molecules are too demanding
and thus far have not been possible in any detail. We have recently
developed a strong-field-approximation theory of HHG by polyatomic
molecules \cite{SenadPolyHHG2016} and illustrated it by the examples
of the ozone and the carbon dioxide molecules exposed to a
linearly polarized laser field. In the present work we apply this
theory to more complex molecules and more complex laser fields. Our
goal is to relate the respective molecular symmetry to the symmetry of the laser field and
to explore the influence of these symmetries on the HHG spectra. We
neglect the vibrational dynamics during the HHG process since it has
recently been shown that for a broad range of molecules in strong
ultrashort laser pulses it does not play an important role
\cite{SandorPRL2016}.

A special field combination, called the bicircular field, has
recently attracted a lot of attention. A bicircular field consists of
two coplanar counterrotating circularly polarized fields with
frequencies $r\omega$ and $s\omega$, which are integer multiples of
the fundamental frequency $\omega$. In our notation this field is
defined in the $xy$ plane by 
\cite{Odzak2015}
\begin{eqnarray}
E_x(t)&=&\left[E_1\sin(r\omega t)+E_2\sin(s\omega t)\right]/\sqrt{2},\nonumber\\
E_y(t)&=&\left[-E_1\cos(r\omega t)+E_2\cos(s\omega
t)\right]/\sqrt{2}.\label{FieldComp}
\end{eqnarray}
A possible phase offset of the first component is canceled by an appropriate
choice of the initial time, while an offset the phase of the second component
is canceled by a rotation of the field around the $z$ axis by the angle
$-\phi_2/(r+s)$ (for $s\omega t\rightarrow s\omega t+\phi_2$).

HHG by such field has been shown experimentally to be very efficient as early
as in 1995 \cite{bicir1995} and was investigated theoretically in
subsequent years
\cite{bicir1995T,JNOPM95,Alon1998,bicir1999,bicir2000,bicir2000RC,pisa,Bauer2001,Nilsen2002,BL03}.
Crucial for the revival of interest in strong-field processes in bicircular
fields was the experimental confirmation that the generated harmonics are
circularly polarized \cite{Fleischer2014}. This paper has triggered
a series of papers about HHG by atoms exposed to bicircular fields
\cite{Fleischer2014_1,Pisanty2014,Kfir2015,OL2015,JPBFTC2015,MedPRL2015,MiloPRA2015,PNAS,ScienceAdvances,MadsenPRA2016}.
Other atomic processes in strong bicircular fields were also
investigated, such as strong-field ionization
\cite{bicirATD,bicirATI,BicirHATIOE,bicirATI1,bicirATI2},
laser-assisted recombination \cite{Odzak2015}, nonsequential double
ionization \cite{NSDI}, the possibility of introducing spin into
attoscience \cite{spin}, etc. Molecular HHG using a bicircular field
was studied in \cite{MolBicir,UzerJPB2016,CPB2016}.

The electric field vector of the $r\omega$--$s\omega$ bicircular
field is invariant with respect to simultaneous translation in time
by $pT/(r+s)$, where $p$ is an integer and $T=2\pi/\omega$ is the
optical period, and rotation about the $z$ axis, which is
perpendicular to the polarization plane, by the angle
$pr2\pi/(r+s)$. For the $\omega$--$2\omega$ field this dynamical
symmetry means that $R_z(p2\pi/3)\vE(t+pT/3)=\vE(t)$, i.e. for $p=1$
the translation in time is by one third of the optical cycle and the
rotation is by the angle $120^\circ$. If the laser-free molecular
Hamiltonian is invariant under rotation by the angle $pr2\pi/(r+s)$
then it can be shown in the same way as in Appendix A of
Ref.~\cite{MiloPRA2015} that the afore-mentioned dynamical symmetry leads
to the following selection rule: only harmonics of the order
\begin{equation}
n=q(r+s)\pm r, q - \mathrm{integer}, \label{S11new}
\end{equation}
are emitted and they are circularly polarized with the helicity $\varepsilon_n=\pm 1$, respectively. If a molecule possesses the rotational symmetry $C_{r+s}$ then the
Hamiltonian of this molecule is invariant with respect to a
rotation by the angle $2\pi/(r+s)$ about the $z$ axis, which is
chosen to be the molecular quantization axis
\cite{Levine,Schonland}. For example, the molecules NH$_3$, CF$_3$I,
SO$_3$, BH$_3$, and BF$_3$ possess the $C_3$ symmetry so that the
Hamiltonians of these molecules are invariant with respect to
rotation by the angle $p\cdot 120^\circ$. If these molecules are
exposed to the $\omega$--$2\omega$ bicircular field in the
$xy$ polarization plane, the harmonics $n=3q$,
$q=1,2,3,\ldots$, are absent from their HHG spectra,
while the harmonics $n=3q\pm 1$ are
circularly polarized with the ellipticity $\varepsilon_n=\pm 1$.

The power of the $n$th harmonic having the wave vector $\vK$ and the
unit complex polarization vector $\veK$ is
\cite{SenadPolyHHG2016,MiloPRA2015}
\begin{equation}
P_n=\frac{(n\omega)^4}{2\pi c^3}\left|\vT_n\right|^2, \;\;
\vT_n=\int_0^T\frac{dt}{T}e^{in\omega t}\sum_m\vd_m(t),\label{T1}
\end{equation}
where $\vT_n =T_n\veK=\sum_{\xi=x,y,z}T_{n\xi}\ve_{\xi}$ and
$\vd_m(t)$ is the time-dependent dipole which corresponds to the
$m$th molecular orbital of the neutral polyatomic molecule having closed
shells. Within the strong-field approximation, the time-dependent
dipole is given by \cite{SenadPolyHHG2016}
\begin{eqnarray}\label{dfi_final}
\vd_m(t)&=& -i\left(\frac{2\pi}{i}\right)^{3/2}\int_0^\infty
\frac{d\tau}{\tau^{3/2}}e^{-iS_s}\nonumber\\
&&\times\sum_{k=1}^N e^{i\left[\vk_s+\vA(t)\right]\cdot\ver_k}
\langle\varphi_{km}|\ver|\vk_s+\vA(t)\rangle\\
&&\times\sum_{j=1}^N e^{-i\vk_s\cdot\ver_j}
\langle\vk_s+\vA(t-\tau)|\ver\cdot\vE(t-\tau)|\varphi_{jm}\rangle,\nonumber
\end{eqnarray}
with $S_s\equiv \int_{t-\tau}^tdt'[\vk_s+\vA(t')]^2/2+\IP\tau$,
$\IP$ the ionization potential,
$\vk_s=-\int_{t-\tau}^tdt'\vA(t')/\tau$, and $\vE(t)=-d\vA(t)/dt$.
The molecular orbitals
$\phi_m(\ver)=\sum_{j=1}^N\varphi_{jm}(\ver)=\sum_{j=1}^N
\sum_{a}c_{jam}\psi_{am}(\ver)$ are written as linear combinations
of the atomic orbitals $\psi_{am}$, which are expressed as linear
combinations of primitive Gaussian functions \cite{Levine}. Here
$\ver=\ver_e-\ver_j$, where $\ver_j$, $j=1,\ldots, N$, are the
position vectors of the atoms and $\ver_e$ is the electron
coordinate.

\begin{figure}
\begin{center}
\includegraphics[width=4.5cm]{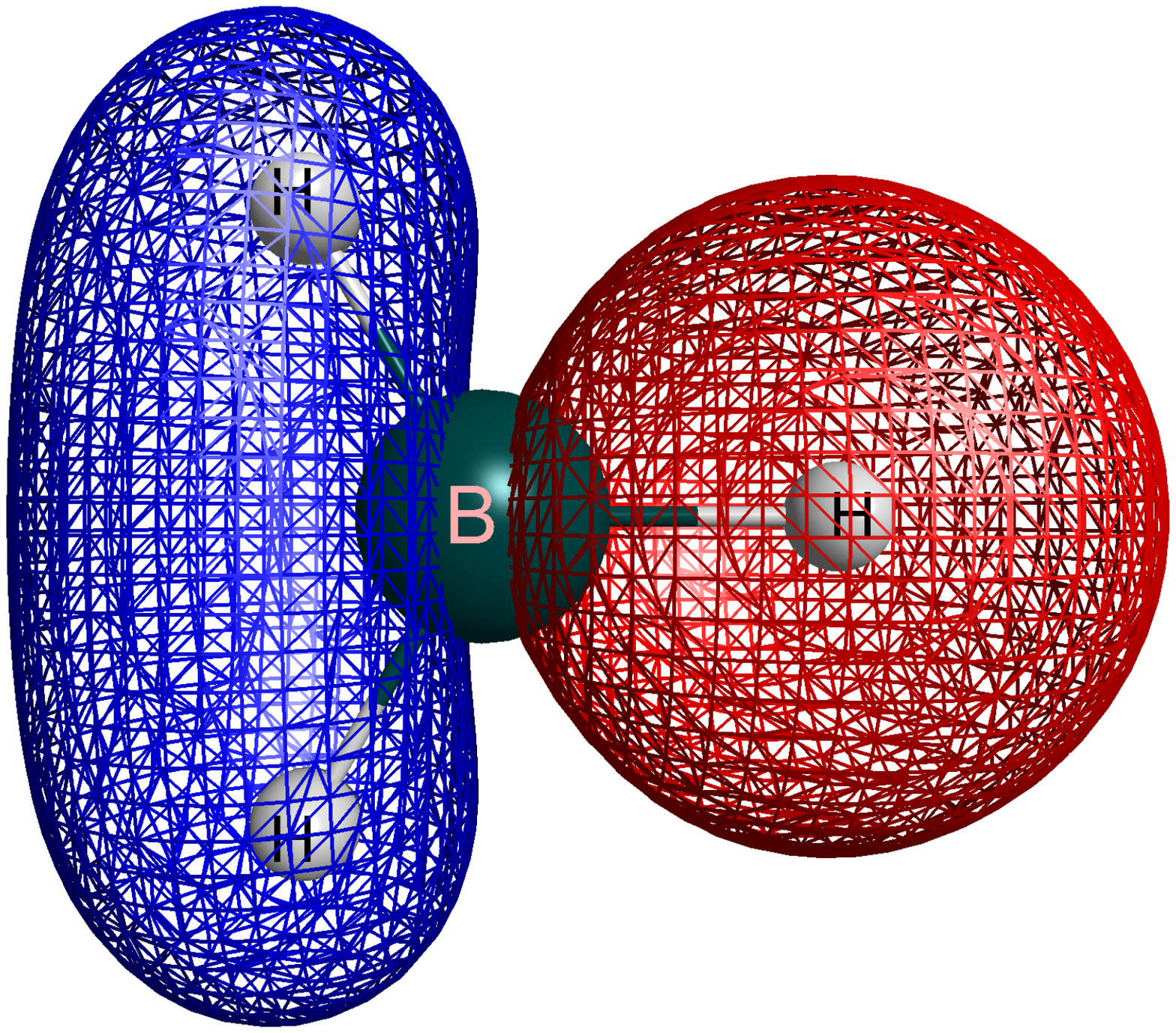}\includegraphics[width=4.5cm]{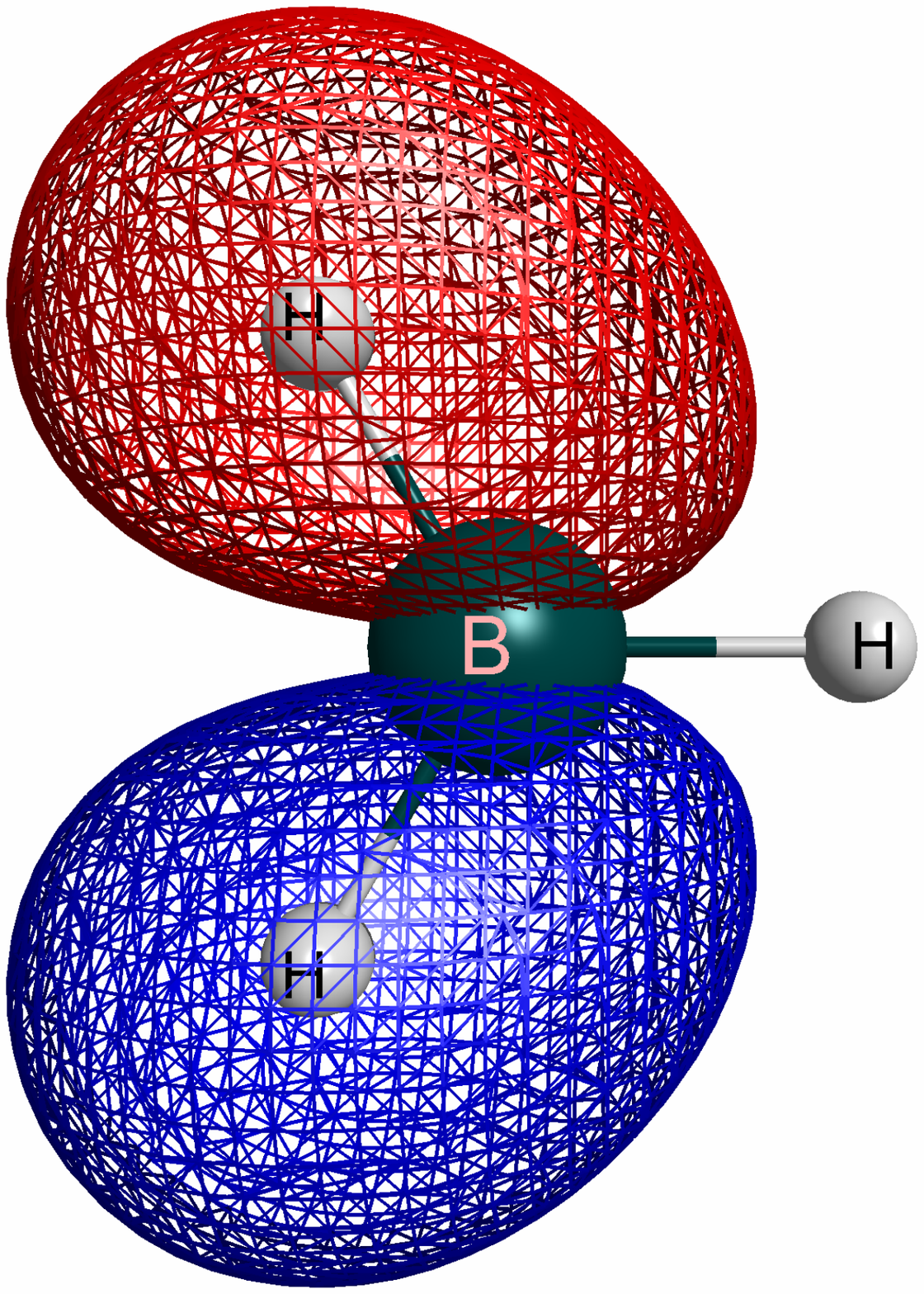}\\
\includegraphics[width=8cm]{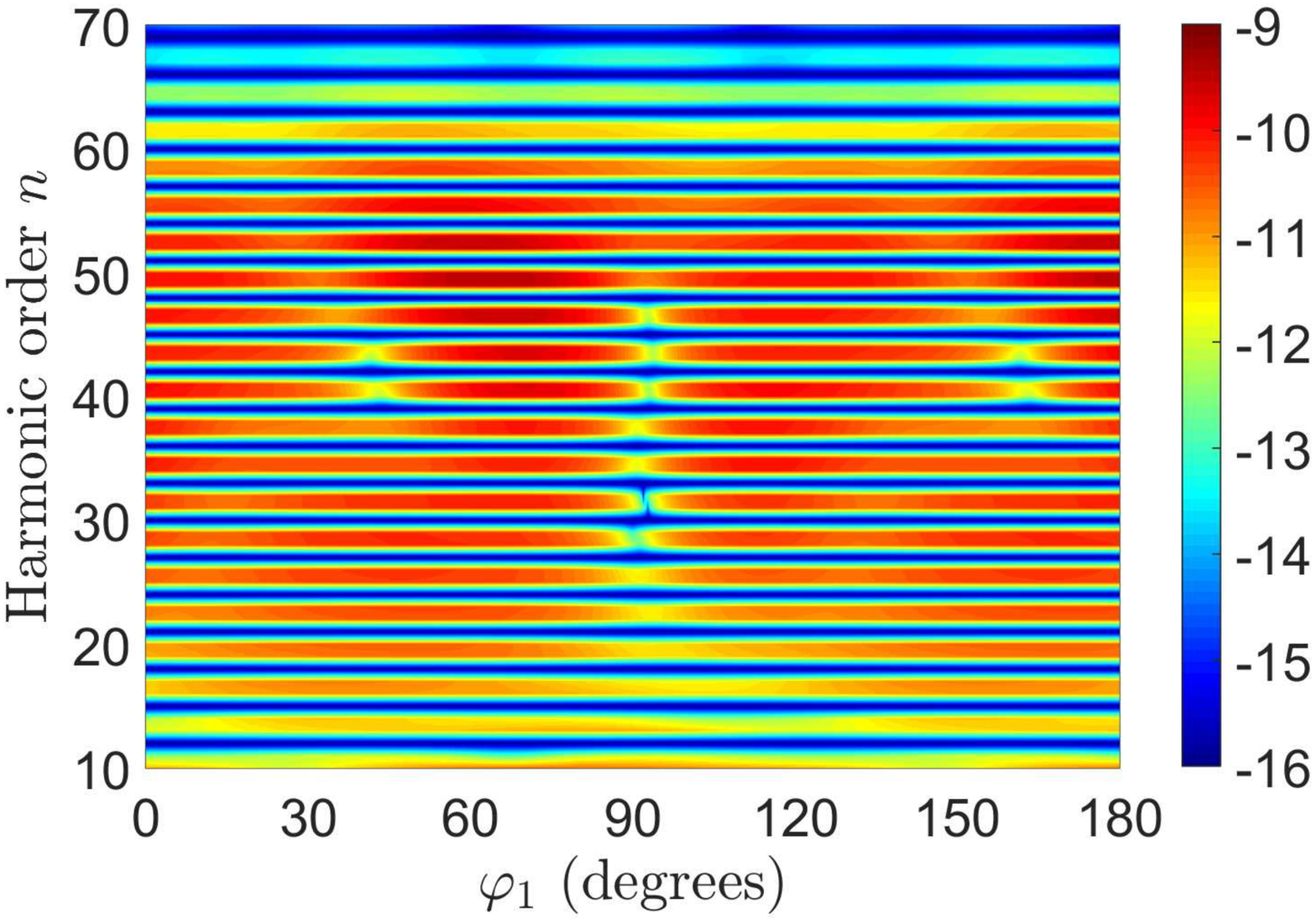}\\
\includegraphics[width=8cm]{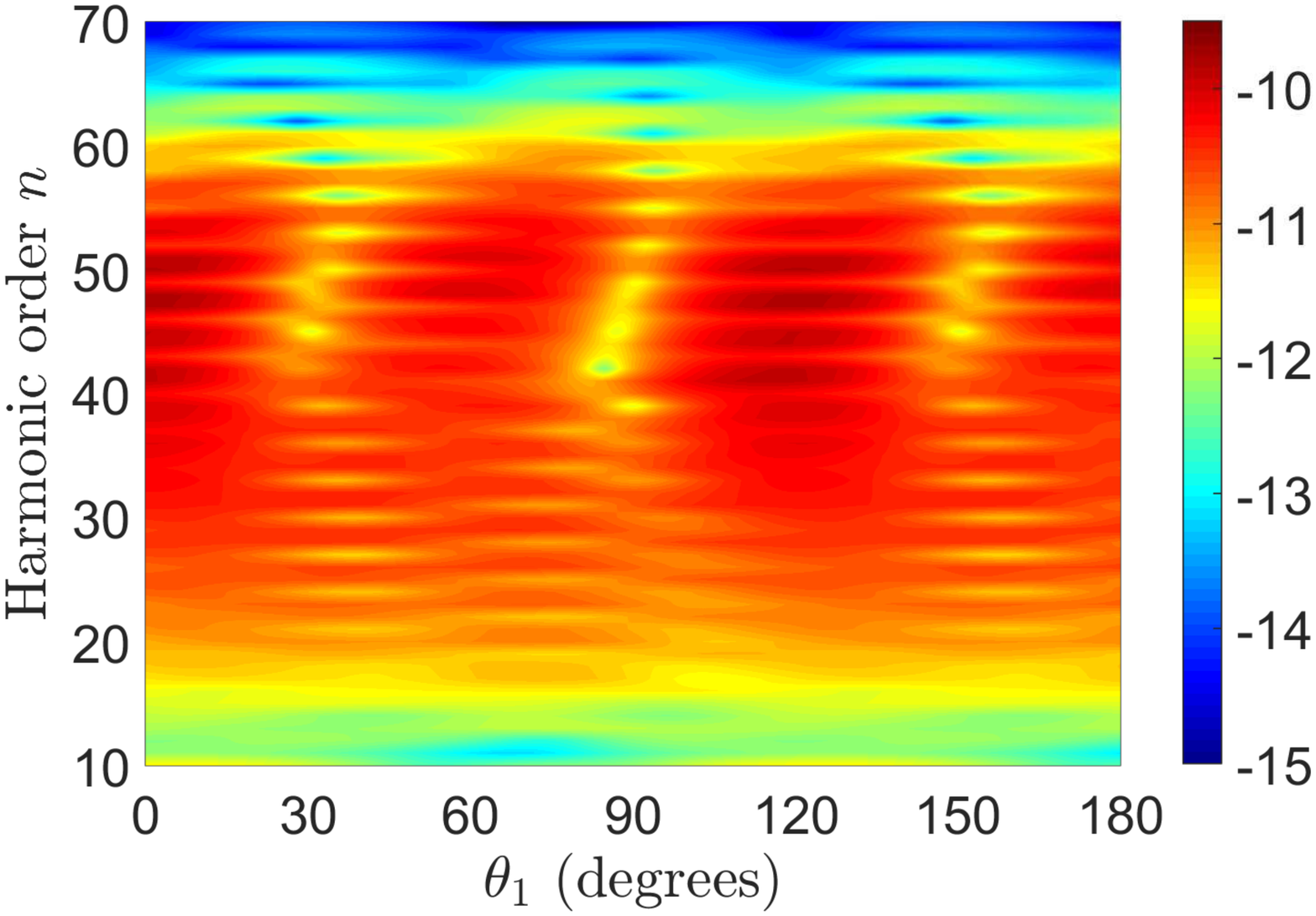}
\end{center}
\caption{(Color online) Upper panel: two degenerate HOMOs of the BH$_3$
molecule in the $xy$ plane. Middle (lower) panel: logarithm of the
harmonic power for the BH$_3$ molecule presented in false colors as a
function of the rotation angle $\varphi_1$ ($\theta_1$) in the $xy$
($xz$) plane (see Supplementary information for the definition of
this rotation). The bicircular laser field component intensities are equal
$4\times 10^{14}\Wcm$ and the fundamental wavelength is
$800$~nm.}\label{fig:BH3}
\end{figure}
For our calculation we will use the examples of the planar BH$_3$
and BF$_3$ molecules (see Supplementary information), which both
possess the $C_3$ symmetry. The highest occupied molecular orbital
(HOMO) of the BH$_3$ molecule is doubly degenerate having the
ionization potential $\IP=13.59$~eV and the $E'$ symmetry. The
ionization energies of the other molecular orbitals are much higher
and their contributions can be neglected. Therefore, in
Eq.~(\ref{dfi_final}) we have the sum over two degenerate molecular
orbitals. Even though the molecular orbitals are not symmetric with
respect to rotation by the angle $120^\circ$ about the $z$ axis (see
the upper panel of Fig.~\ref{fig:BH3}), the HHG spectra for the
bicircular $\omega$--$2\omega$ field in the molecular $xy$ plane
(middle panel of Fig.~\ref{fig:BH3}) show that the harmonics $n=3q$
are absent. On the other hand, if the bicircular field is not in the
molecular plane the symmetry is violated and all harmonics are
emitted. This is clearly visible in the bottom panel of
Fig.~\ref{fig:BH3} where the bicircular field rotates in the $xz$
plane.

\begin{figure}
\begin{center}
\includegraphics[width=7cm]{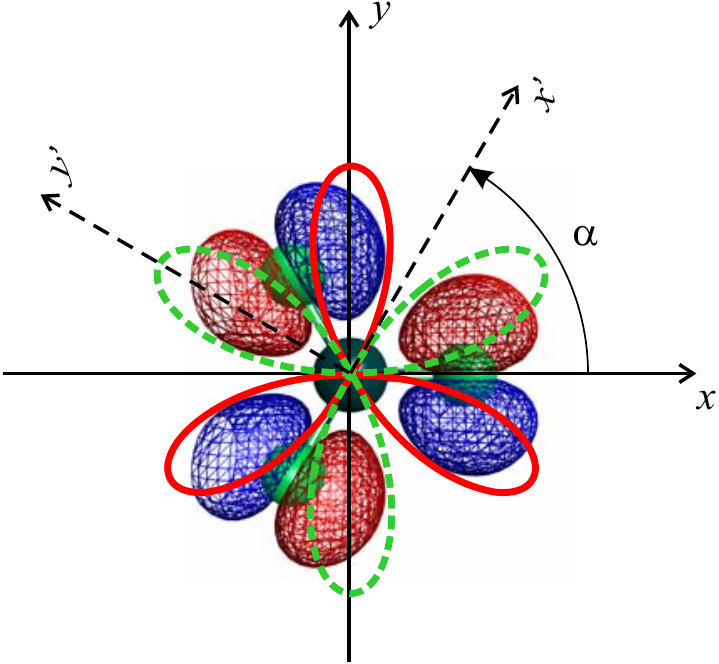}
\end{center}
\caption{(Color online) Bicircular field and its rotation in the
$xy$ plane by the angle $\alpha=60^\circ$, presented together with
the HOMO of the BF$_3$ molecule.}\label{fig:2}
\end{figure}
The BH$_3$ molecule is relatively simple and no particular structure
is visible in the spectra presented in Fig.~\ref{fig:BH3}. Let us
now consider the BF$_3$ molecule, which possesses the same $C_3$
symmetry but the H atoms are replaced by more complex F atoms. The
orbitals of the BF$_3$ molecule are closer in energy (see
Table~\ref{tab:II} in Supplementary Information). The orbitals
denoted by a double prime ($''$) are antisymmetric with respect to
reflection \cite{Levine} about the horizontal ($xy$) plane so that the
corresponding ionization matrix element in Eq.~(\ref{dfi_final}) is
zero (for $z\rightarrow -z$ the field does not change). Therefore,
HOMO-1 and HOMO-3, having, respectively, $E''$ and $A_2''$ symmetry,
do not contribute to the harmonic emission. The HOMO of the BF$_3$
molecule, having the $A_2'$ symmetry \cite{Levine}, is presented in
Fig.~\ref{fig:2}, together with the bicircular field, which is in the
$xy$ plane and rotated by the angle $\alpha=\varphi_1=0^\circ$ (red
solid line) and $60^\circ$ (green dashed line).

\begin{figure}
\begin{center}
\includegraphics[width=8cm]{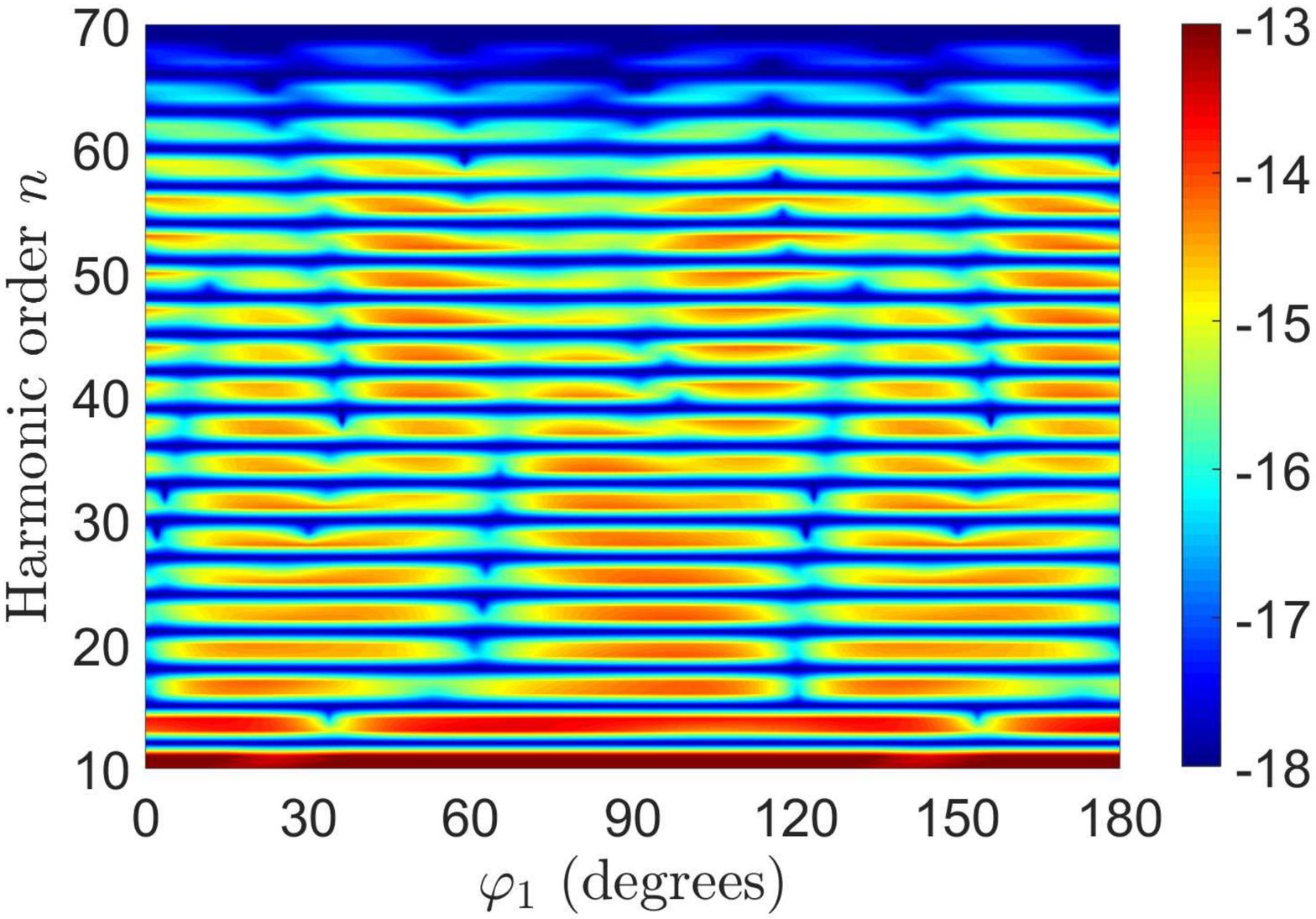}\\
\includegraphics[width=8cm]{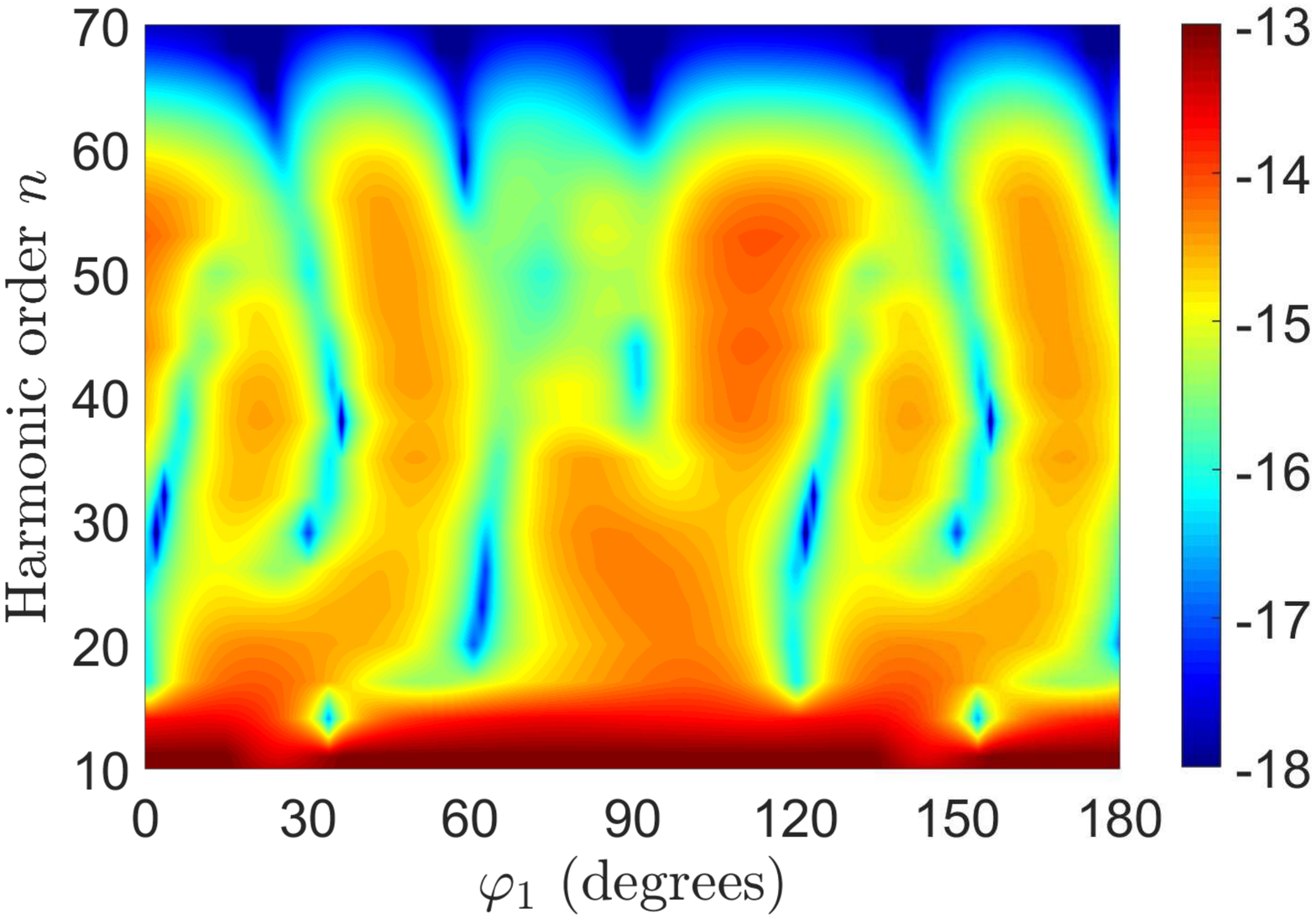}\\
\includegraphics[width=8cm]{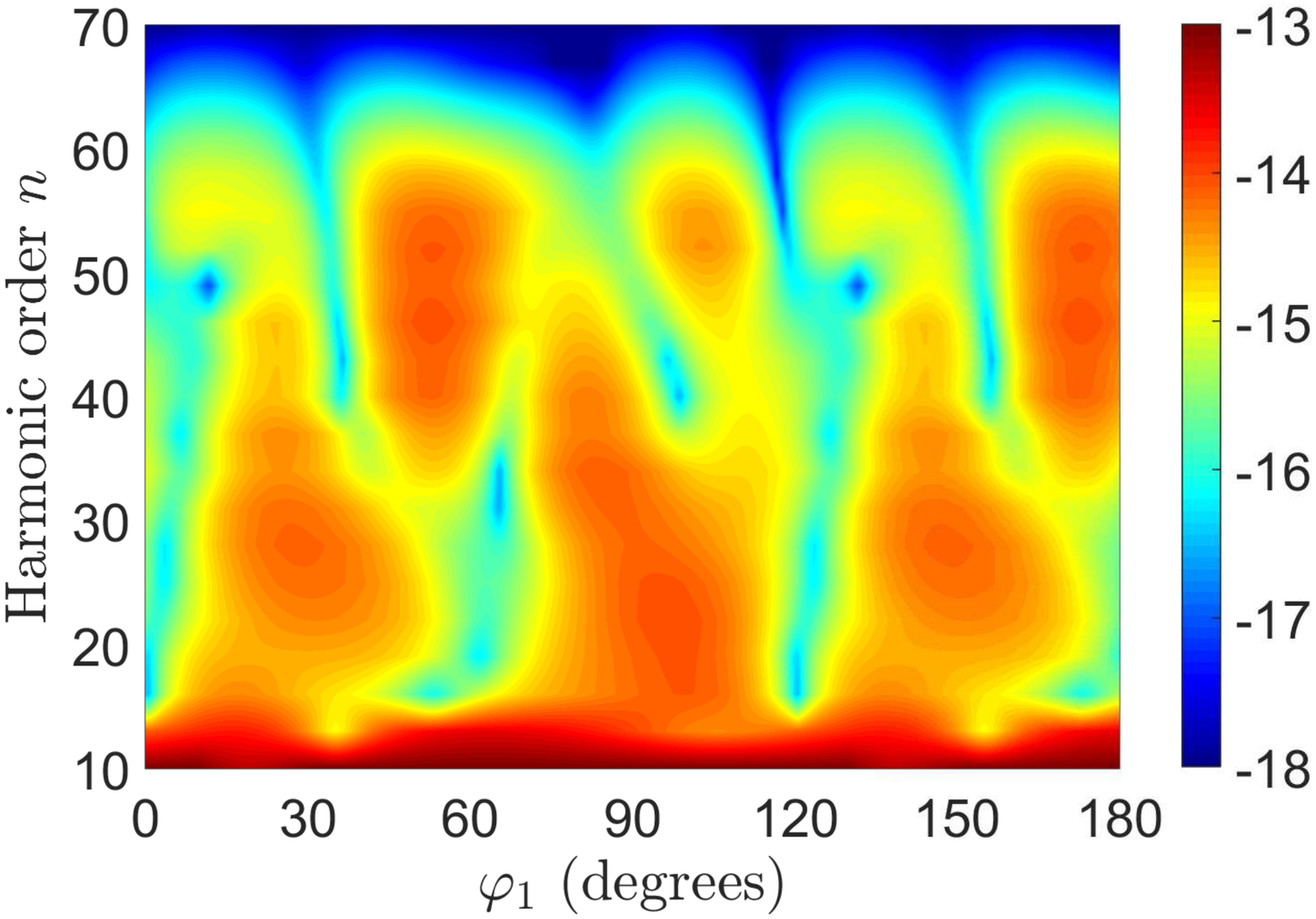}
\end{center}
\caption{(Color online) Upper panel: logarithm of the harmonic power
for the BF$_3$ molecule presented in false colors as a function of the
rotation angle $\varphi_1$  in the $xy$ plane (horizontal axis) and
harmonic order $n$ (vertical axis). Middle (bottom) panel: same as
in the upper panel but with only the harmonics having ellipticity
$\varepsilon_n=-1$ ($+1$). The parameters of  the bicircular laser field
 are as
in Fig.~\ref{fig:BH3}.}\label{fig:BF3}
\end{figure}
In Fig.~\ref{fig:BF3} we present the HHG spectra for the BF$_3$ molecule
and the $\omega$--$2\omega$ bicircular field in the $xy$
polarization plane. From the upper panel of Fig.~\ref{fig:BF3} we
see that only the harmonics $n=3q\pm 1$ are emitted for arbitrary
rotation angle $\varphi_1$ in the $xy$ plane. In addition, more
structures are visible in the spectrum. These structures are
invariant with respect to translation by $120^\circ$ (the results
for $0\le\varphi_1\le 60^\circ$ are identical to the results for
$120\le\varphi_1\le 180^\circ$). The interference structure visible
in the spectrum can be related to the shape of the HOMO. From
Fig.~\ref{fig:2} we see that the polar diagram of the bicircular
electric field vector for $\alpha=0^\circ$ partially overlaps with
the ``blue'' part of the electron density, while for
$\alpha=60^\circ$ it overlaps with the ``red'' part for which the
electron wave function has the opposite sign. For different $\alpha$
this overlap is different which is the cause of the change of the
interference structure.

\begin{figure}
\begin{center}
\includegraphics[width=7.5cm]{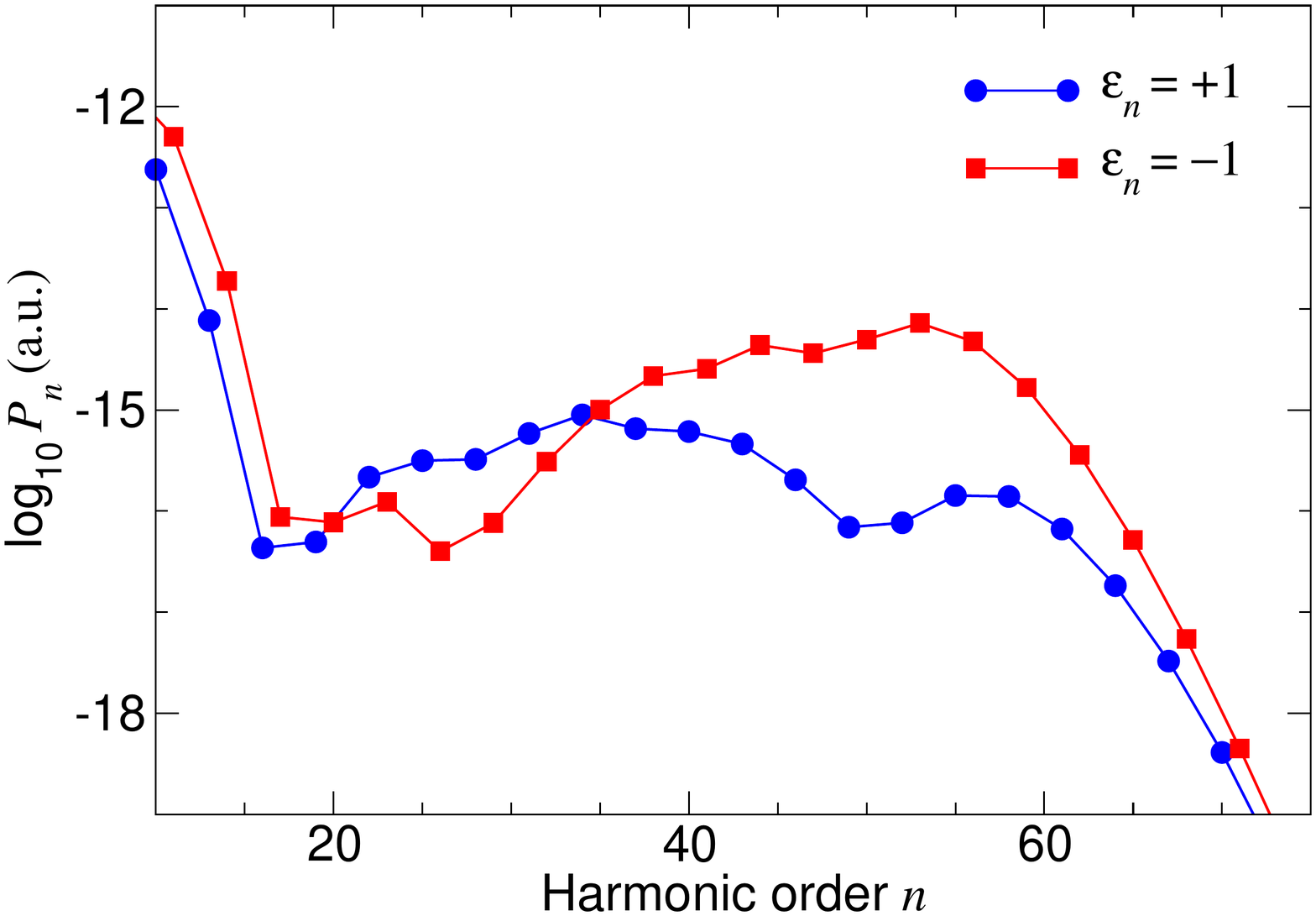}\\
\vspace{1cm}\includegraphics[width=7.5cm]{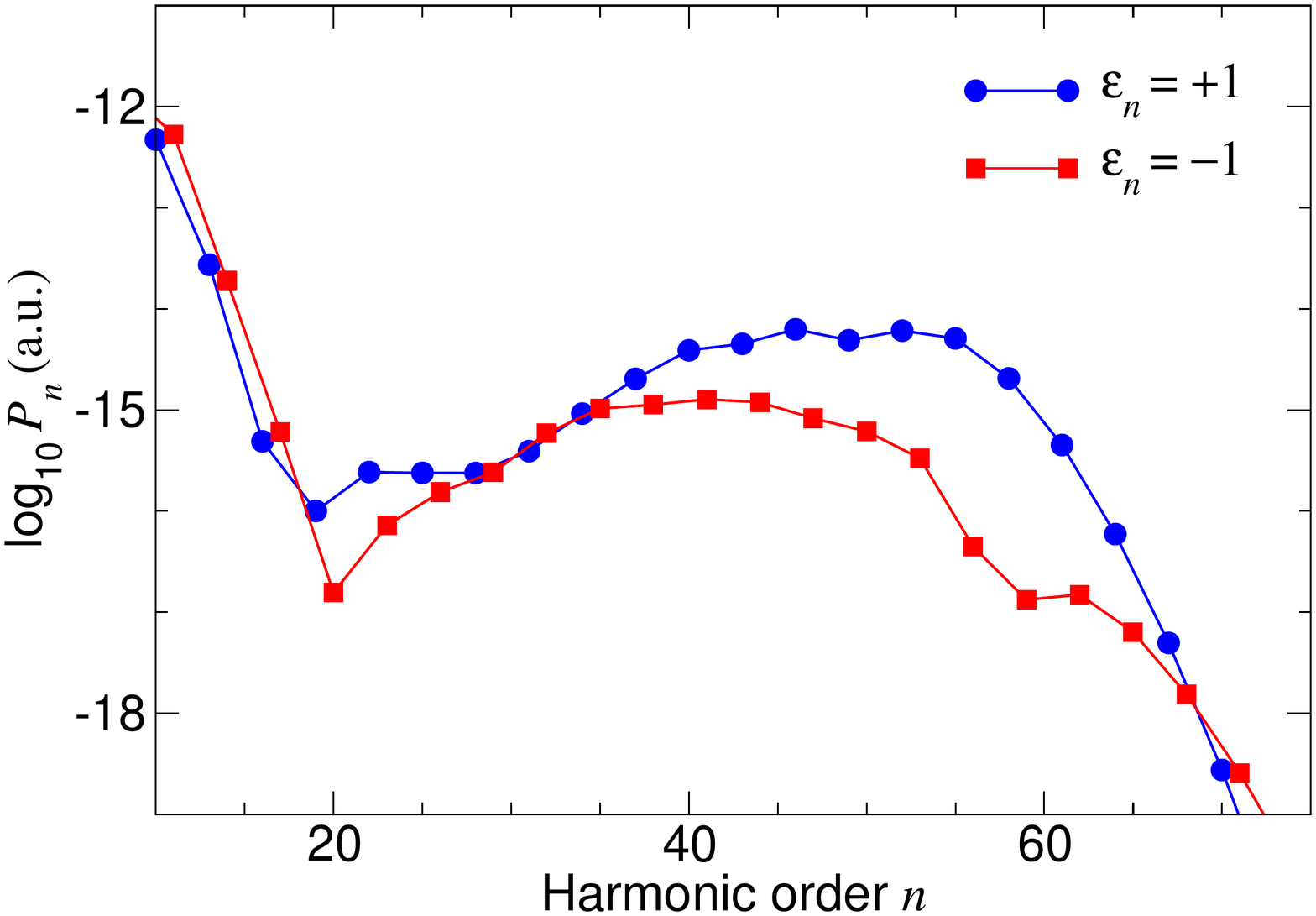}
\end{center}
\caption{(Color online) Logarithm of the harmonic power for the BF$_3$
molecule presented as a function of the harmonic order for the same
laser parameters as in Fig.~\ref{fig:BH3}, for the HOMO and for
$\varphi_1=0^\circ$ (upper panel) and $\varphi_1=60^\circ$ (lower
panel). The bicircular field is in the $xy$ plane as denoted in the
upper panel of Fig.~\ref{fig:2}. The harmonic ellipticity is
indicated in the legend. }\label{fig:3}
\end{figure}
It has been shown in Refs.~\cite{OL2015,MiloPRA2015} that for atomic
HHG from a $p$ ground state the yields of the  $3q+1$ and $3q-1$
harmonics are different. Since the F atoms in the BF$_3$ molecule
behave as $p$-state atoms we expect that such an asymmetry will also
appear for the BF$_3$ molecule. This is clearly visible in the
middle and bottom panels of Fig.~\ref{fig:BF3}, where the spectra of
harmonics $n=3q-1$ with $\varepsilon_n=-1$ (middle panel) and the
harmonics $n=3q+1$ with $\varepsilon_n=+1$ (bottom panel) are shown.
The results are interpolated and smoothed for a better visibility of
the interference structures. As a further example, in
Fig.~\ref{fig:3} we compare the $\varepsilon_n=+1$ and
$\varepsilon_n=-1$ spectra for two fixed values of the angle
$\varphi_1=\alpha=0^\circ$ (upper panel) and $60^\circ$ (lower
panel), which correspond to the fields presented in
Fig.~\ref{fig:2}. The mentioned asymmetry is clearly visible in the
plateau region. The $\varepsilon_n=-1$ harmonics are dominant for
$\varphi_1=0^\circ$, while for $\varphi_1=60^\circ$ the situation is
opposite and the $\varepsilon_n=+1$ harmonics are dominant.
Comparing the middle and bottom panels of Fig.~\ref{fig:BF3} we can
find more regions with such asymmetry. For example, for harmonic
order near 30th and the angle $\varphi_1\approx 30^\circ$ there is a
large dip in the harmonic power for $\varepsilon_n=-1$ (middle
panel), while for $\varepsilon_n=+1$ (bottom panel) there is a
strong maximum. By inspecting Fig.~\ref{fig:2} one can relate this
change in the asymmetry with the above-discussed overlap of the
electron density and the electric field.

How we can use the results of the present work for future
experiments? First, the absence of the $n=3q$ harmonics from the HHG
spectra of planar molecules having $C_3$ symmetry can be used as a
measure of the orientation of the molecules in the plane. The
measured parameter can be the ratio of the harmonic power
$P_{3q}/P_{3q\pm 1}$ (it is zero for perfectly oriented molecules).
Second, for molecules which are already oriented in a plane one can
record the harmonic power of harmonics having different helicities
as a function of the angle $\varphi_1$. The rotation angle of the
bicircular field can easily be controlled in an experiment by
changing the relative phase of the field components (see, for
example, the experimental realization in Ref.~\cite{bicirATI}). From
these spectra one can extract information about the dominant
molecular orbitals. In Figs.~\ref{fig:BH3} and \ref{fig:BF3} the
results for the HOMOs are presented. Similar calculations can be
done for other molecules and different molecular orbitals and be used
for comparison with experiments.

In conclusion, high harmonics emitted by polyatomic molecules that
obey the $C_{r+s}$ symmetry, exposed to an $r\omega$--$s\omega$
bicircular field with the polarization plane perpendicular to the
molecular quantization axis, can only have the order $n=q(r+s)\pm r$
and the corresponding ellipticity $\varepsilon_n=\pm 1$. We
illustrated this by the examples of the planar molecules BH$_3$ and
BF$_3$, which obey the $C_3$ symmetry. For the BF$_3$ molecule,
similarly to atoms having a $p$ ground state, there is a
strong asymmetry in the emission of high harmonics of opposite
helicities. This asymmetry depends on the molecular orientation and
may be used for its determination. As in
Refs.~\cite{OL2015,MedPRL2015,MiloPRA2015} for atoms, this asymmetry
can be used to generate isolated elliptically polarized attosecond
pulses by polyatomic molecules.

\begin{acknowledgments}
We thank Wilhelm Becker for useful comments. Support by the Federal
Ministry of Education and Science, Bosnia and Herzegovina is
gratefully acknowledged.
\end{acknowledgments}

\appendix
\section{Supplementary information}\label{sec:suppl}
\subsection{Molecular characteristics}\label{subsec:molchar}
In this Supplementary Information, we show the equilibrium
geometries, symmetries, and energies of the molecular orbitals for
BH$_3$ and BF$_3$ molecules. The presented results are calculated
using the Hartree-Fock method with cc-pVTZ basis set from the GAMESS
quantum chemistry package \cite{SchmidtJCC1993}.

The BH$_3$ and BF$_3$ molecules correspond to the group $D_{3h}$
which means that they have a $C_3$ axis, three $C_2$ axes and a
$\sigma_h$ symmetry plane perpendicular to the $C_3$ axis. They also
have three vertical planes of symmetry, each such plane passing
through the $C_3$ axis and a $C_2$ axis. The results for the
equilibrium geometries are shown in Table~\ref{tab:I}, while the
symmetries and energies of the molecular orbitals are presented in
Table~\ref{tab:II}.

\begin{table}
\caption{The equilibrium geometries for BH$_3$ and BF$_3$ molecules
calculated using the GAMESS \cite{SchmidtJCC1993}.\\}\label{tab:I}
\begin{center}
\begin{tabular}{|c|c|c|c|}
\hline
 atom & $x$ (\AA) & $y$ (\AA) & $z$ (\AA) \\
 \hline
 B & 0.0000 & 0.0000 & 0.0000\\
 \hline
 H & -0.5937 & 1.0283 & 0.0000\\
 \hline
 H & -0.5937 & -1.0283 & 0.0000\\
 \hline
 H & 1.1874 & 0.0000 & 0.0000\\
 \hline
\end{tabular}
\end{center}
\begin{center}
\begin{tabular}{|c|c|c|c|}
\hline
 atom & $x$ (\AA) & $y$ (\AA) & $z$ (\AA) \\
 \hline
 B & 0.0000 & 0.0000 & 0.0000\\
 \hline
 F & -0.6473 & 1.1212 & 0.0000\\
 \hline
 F & -0.6473 & -1.1212 & 0.0000\\
 \hline
 F & 1.2947 & 0.0000 & 0.0000\\
 \hline
\end{tabular}
\end{center}
\end{table}

\begin{table}
\caption{Symmetries and energies of the BH$_3$ and BF$_3$ molecular
orbitals. The cases of two degenerate orbitals are denoted by HOMOs,
HOMO-1s etc.
\\}\label{tab:II}
\begin{center}
\begin{tabular}{|c|c|c|}
\hline
BH$_3$ orbital & symmetry & energy (eV) \\
 \hline
HOMO-1 & $A_1'$ & $-19.19$ \\
 \hline
HOMOs & $E'$ & $-13.58$ \\
 \hline
\end{tabular}
\end{center}
\begin{center}
\begin{tabular}{|c|c|c|}
\hline
BF$_3$ orbital & symmetry & energy (eV) \\
 \hline
HOMO-6s & $E'$ & $-44.83$ \\
 \hline
HOMO-5 & $A_1'$ & $-23.40$ \\
 \hline
HOMO-4s & $E'$ & $-22.24$ \\
 \hline
HOMO-3 & $A_2''$ & $-20.88$ \\
 \hline
HOMO-2s & $E'$ & $-19.08$ \\
 \hline
HOMO-1s & $E''$ & $-18.69$ \\
 \hline
HOMO & $A_2'$ & $-17.99$ \\
 \hline
\end{tabular}
\end{center}
\end{table}

\subsection{Definition of rotation} \label{subsec:rot} Molecules are
defined in a fixed Cartesian $xyz$ coordinate system. Passive
rotation by the Euler angles of this system makes the $x$ axis in
the direction $\ve_1$ ($x$ axis of the bicircular field) and the $y$
axis in the direction $\ve_2$ ($y$ axis of the bicircular field).
According to \cite{Rose}, the product of three Euler rotations is
$M(\alpha\beta\gamma)=M(\gamma)M(\beta)M(\alpha)$, where $\alpha$ denotes
rotation about the original $z$ axis, $\beta$ about the new $y$
axis, and $\gamma$ about the final $z$ axis. In accordance with
the above definition we have $ M(\alpha\beta\gamma) (1,0,0)^T
=\ve_1=(\sin\theta_1\cos\varphi_1,\sin\theta_1\sin\varphi_1,
\cos\theta_1)^T$, $M(\alpha\beta\gamma) (0,1,0)^T
=\ve_2=(\sin\theta_2\cos\varphi_2,\sin\theta_2\sin\varphi_2,
\cos\theta_2)^T$ (the superscript $T$ indicates the transpose matrix), where
the angles $(\theta_1,\varphi_1)$ and $(\theta_2,\varphi_2)$,
respectively, define the vectors $\ve_1$ and $\ve_2$ in the
spherical coordinates of the molecular system.

If the bicircular field polarization plane is the $xy$ plane then we
have  $\varphi_1=\varphi_2-90^\circ=\alpha$ and
$\theta_1=\theta_2=90^\circ$, $\beta=\gamma=0^\circ$. For the field
in the $xz$ plane we have $\alpha=-\theta_1$,
$\beta=\gamma=90^\circ$. In this case for
$\theta_1\in[0^\circ,90^\circ]$ we have
$\varphi_1=\varphi_2=0^\circ$, $\theta_2=\theta_1+90^\circ$, while
for $\theta_1\in[90^\circ,180^\circ]$ it is $\varphi_1=0$,
$\varphi_2=180^\circ$, $\theta_2=270^\circ-\theta_1$.

\end{document}